\documentclass[prd,aps,showpacs]{revtex4}

\usepackage{epsfig}
\usepackage{amsmath,amsthm,amssymb}
\usepackage{bm}

\begin{document}

\title{Dwarf Galaxies, MOND, and Relativistic Gravitation}

\date{\today}
\author{Arthur Kosowsky}
\email{kosowsky@pitt.edu}
\affiliation{Department of Physics and
Astronomy, University of Pittsburgh, 3941 O'Hara Street, Pittsburgh, PA 15260 USA}

\begin{abstract}
MOND is a phenomenological modification of Newton's law of gravitation which
reproduces the dynamics of galaxies, without the need for additional dark matter. 
This paper reviews the basics of MOND and its application to dwarf galaxies. MOND
is generally successful at reproducing stellar velocity dispersions in the Milky Way's classical dwarf ellipticals, for reasonable values of the stellar mass-to-light ratio of the galaxies; two
discrepantly high mass-to-light ratios may be explained by tidal effects. Recent observations also show MOND describes tidal dwarf galaxies, which form in complex dynamical environments. The application of
MOND to galaxy clusters, where it fails to reproduce observed gas temperatures, is also
reviewed. Relativistic theories containing MOND in the non-relativistic limit have now been formulated; they all contain new
dynamical fields, which may serve as additional sources of gravitation that could reconcile cluster observations with MOND. Certain limits of these theories can also give the
accelerating expansion of the Universe. The standard
dark matter cosmology boasts numerous manifest triumphs; however,
alternatives should also be pursued as long as outstanding observational
issues remain unresolved, including the empirical successes of MOND on galaxy scales 
and the phenomenology of dark energy.

\end{abstract}


\maketitle

\section{Why MOND?}

In 1983, Mordehai Milgrom published a set of three papers in the Astrophysical Journal
postulating a fundamental modification of either Newton's Law
of Gravitation or Newton's Second Law \cite{milgrom83a,milgrom83b,milgrom83c}. This modification was designed to explain the Tully-Fisher relation for spiral galaxies, which relates luminosity to rotational velocity, without using dark matter. Unlike many other proposed modifications
of gravity, Milgrom's is not at a particular length or energy scale, but rather at 
an {\it acceleration} scale $a_0 = 1.2\times 10^{-10}$ m/s$^2$. This is a very small
acceleration, which makes any laboratory-based probes nearly impossible; the
consequences of Milgrom's modification manifest themselves only on astronomical
scales. In particular, Milgrom aimed to address the phenomenology of dark matter:
are the departures from Newtonian gravitation we see in galaxies actually due to
some breakdown of Newton's Law of Gravitation, rather than due to the action of
otherwise undetected dark matter obeying the standard gravitational law? 

For spiral galaxies, redshift measurements
reveal how rapidly stars and neutral hydrogen gas in the disk are rotating around the center of the galaxy.
It has been known since the early 1970's that the overwhelming majority of spiral
galaxies have flat rotation curves: the rotational velocity $v(R)$ at a radius $R$ from
the center of the galaxy becomes independent of $R$ past a certain radius. (A small
fraction of galaxies have rotation curves which are still increasing with $R$ at the largest-radius
data point. Only a few exhibit any significant decrease; see, e.g., \cite{delrio04} for an example.) 
This behavior is impossible to
reconcile with the Newtonian gravitational field produced by the visible matter in the galaxy.
At large distances, beyond the majority of the galaxy's visible mass, the Newtonian gravitational field
from the visible matter begins to approximate the field from a point mass, so that rotational
velocities due to the visible matter alone would drop like $v(R)\propto R^{-1/2}$, as with
planetary orbital velocities in the solar system. No known spiral galaxy exhibits this behavior
in its rotation curve, even though many have rotation curves probed well beyond the bulk of the visible matter
via neutral hydrogen clouds. We have only two possible logical explanations: either a large
amount of unseen dark matter is present to increase the gravitational field over the Newtonian
value, or Newton's Law of Gravitation does not hold on the scales of spiral galaxies. 

From the beginning, dark matter was considered as almost certainly the explanation for
galaxy observations, buttressed by candidate dark matter particles which arise naturally
in many extensions of the standard model of particle physics, particularly in supersymmetry.
The dark matter hypothesis has been extremely successful at explaining a wide range of
cosmological observations, including the growth of large-scale structure
and gravitational lensing. Many experimental groups are in hot pursuit of the direct detection
of dark matter particles, under the assumption that they interact via the weak nuclear force.
Large cosmological N-body simulations have converged on detailed predictions for the distribution
of dark matter on scales ranging from superclusters down to dwarf galaxies, and a large
subfield of cosmology is devoted to figuring out the relationship between the theoretical
dark matter distribution and the observed visible matter distribution, using a range of
observations, simulations, and models. Certainly dark matter cosmology has been highly successful
in explaining a wide range of observations and in providing a compelling theoretical framework
for the standard cosmological model. 

Is there a viable alternative? Milgrom considered the following ad hoc modification
to the Newtonian law for gravitational force on a mass $m$ due to another mass $M$ at a distance $r$:
\begin{equation}
F = 
\begin{cases} 
ma_N ,& a_N \gg a_0;\\
m(a_N a_0)^{1/2}, & a_N \ll a_0,
\end{cases}
\label{mond_force_law}
\end{equation}
where $a_N = GM/r^2$ is the usual Newtonian acceleration due to gravity, and some smooth interpolating function between the two limits is assumed. 
If the Newtonian gravitational force on an object results in an acceleration greater than $a_0$, the Milgrom gravitational force is proportional to $r^{-2}$, the same as the Newtonian force, but if the Newtonian force gives an acceleration smaller than $a_0$, then the Milgrom gravitational force is larger than the corresponding Newtonian force, dropping off like $r^{-1}$.
Remarkably, this simple modification is highly successful at reproducing the rotation curves
of spiral galaxies from the visible matter distribution alone, i.e.\ without dark matter, under the
assumption that stars in disk galaxies follow circular orbits around the center of the galaxy. 
Many rotation curves have been fit this way; see, e.g., 
\cite{sanders07,milgrom-sanders07,barnes07,sanders98,deblok98}. Some galaxy rotation curves are not fit by this formula, but these
are generally galaxies which visually appear to be out of dynamical equilibrium or which possess
strong bars or other features departing from circular symmetry, and thus likely violate the
assumption that stars follow circular orbits. Milgrom dubbed this force law modification MOND,
for MOdified Newtonian Dynamics. A large literature has developed about MOND. This paper aims
to give a brief introduction to the current state of the topic with useful points of entry to the literature, but is not a complete review of the literature or the science. 

A cursory analysis of the force law in Eq.~(\ref{mond_force_law}) reveals a fundamental
shortcoming: it does not conserve momentum \cite{felten84}. It is simple to construct a situation with
two different masses where the force of the small mass on the large mass is not the
same as the force of the large mass on the small mass. Shortly after Milgrom's original
papers, Bekenstein and Milgrom proposed a generalization of the Poisson equation,
instead of the Newtonian force law, which overcomes this difficulty \cite{bekenstein84}:
\begin{equation}
\nabla\cdot\left[\mu\left(|\nabla\Phi|/a_0\right)\nabla\Phi\right] = 4\pi G\rho
\label{poisson_equation}
\end{equation}
where $\Phi$ is the physical gravitational potential, $\rho$ is the baryonic mass density,
and $\mu(x)$ is some smooth function with the asymptotic forms $\mu\sim x$, $x\ll 1$ and $\mu\sim 1$, $x\gg 1$. (The precise form of $\mu(x)$ makes little difference in the resulting phenomenology;
commonly $\mu(x)=x/(1+x^2)^{1/2}$ is used. But see also Ref.\ \cite{sanders07}, which claims
that $\mu(x)=x/(1+x)$ gives a more realistic estimate for the visible disk mass in high surface-brightness spirals.) Note that in the regime where $|\nabla\Phi|\gg a_0$ this
equation becomes the usual Poisson equation.
One particular feature of this equation which differs from the usual Newtonian equation is
that it is not linear: the gravitational force on an object cannot be obtained simply by
adding the gravitational forces of all objects acting on it. One implication is the external
field effect: a galaxy will behave according to the MOND force law with the mass of the
galaxy providing the source term only as long as the internal accelerations in the
galaxy are larger than the local acceleration coming from the gravitational field
of external objects. This behavior was postulated in the original Milgrom papers to
explain why open star clusters within our galaxy do not display
MOND-like behavior, and to give a consistent explanation of how to reconcile
the MOND-like behavior of different objects. The modified Poisson Equation in Eq.~(\ref{poisson_equation}) automatically incorporates this behavior, given an
arbitrary mass distribution. 

Milgrom's modified force law not only explains spiral galaxy rotation curves, but also
can explain a number of other phenomenological successes. It predicts that the
amount of inferred dark matter in a  bound system is not a function of the size of the
system, but rather of the characteristic acceleration. This matches what we observe:
the solar system, globular clusters, and some elliptical galaxies appear to have little or no dark matter, while
dwarf galaxies and large spirals have large amounts of dark matter. But the characteristic
acceleration scale of these systems successfully sorts them by dark matter content. The Milgrom
force law also contains the observed Tully-Fisher Relation (luminosity scaling like the fourth power of the asymptotic rotation velocity) in spiral galaxies as a special case, implies Freeman's Law
for spirals (upper limit on surface brightness), and implies
the Fish Law (constant surface brightness) and the Faber-Jackson relation (luminosity scaling with the fourth power of the velocity dispersion) in elliptical galaxies. Milgrom also predicted that
low surface brightness galaxies should be dark matter dominated at all radii, 
which was later confirmed. See Ref.~\cite{sanders02} for an excellent review discussing 
the range of MOND phenomenology.

What is the effectiveness of the Milgrom force law telling us? In his original papers, Milgrom
advocated that the force law is the actual fundamental force law of nature, suggesting that it
either shows a modification of the law of gravity in the weak gravity regime, or a modification
of Newton's second law connecting force and acceleration. Despite observational successes,
most physicists dismiss this possibility. But regardless of its fundamental status, the
Milgrom force law can be viewed as an effective force law for gravitation in galaxies: if
dark matter is causing the excess forces we observe, it must arrange itself in a way
so that Eq.~(\ref{mond_force_law}) represents the total gravitational force due to
the combined visible and dark matter. It is not at all clear in the standard cosmological
model why this should be. (Unfortunately, the immediate assertion in the original
Milgrom papers that Eq.~(\ref{mond_force_law}) represents a fundamental modification of physics caused many cosmologists to ignore the whole business. All cold dark matter advocates
should still have the Milgrom Relation as part of their suite of known facts about galaxies,
like the Tully-Fisher Relation; requiring dark matter models to reproduce this phenomenology
might give interesting clues about the nature of the dark matter or galaxy dynamics.)

\section{MOND and Dwarf Galaxies}

As dwarf galaxies (particularly dwarf spheroidals) 
exhibit the largest discrepancies between visible matter and gravitational field of any
known bound systems,
they are important laboratories for testing both the dark matter and the modified gravity hypotheses. 
On the other hand, local dwarfs are usually irregular or spheroidal, and thus do not present an easily-measured
rotation curve. Generally, redshift measurements of individual stars in dwarf spheroidals give
an estimate of velocity dispersion as a function of the angular distance from the dwarf center,
along with the dwarf's overall line-of-sight velocity.
These measurements provide two possible probes of MOND: are internal dynamics of dwarf galaxies 
consistent
with the MOND force law, and are the motions of dwarf galaxies around their parent galaxies consistent with
the MOND force law? We briefly consider both questions here.

\subsection{Internal Dynamics of Dwarfs}

MOND was originally formulated to fit measured rotation curves of spiral galaxies. Extrapolating
from the force law in Eq.~(\ref{mond_force_law}) led to the prediction that low surface-brightness galaxies should have rotation curves which implied dark matter domination at all radii from the center; this prediction was eventually born out by later observations. Extending this
analysis to dwarf spheroidal galaxies, which also have low surface brightness,
Milgrom also predicted ``...when velocity-dispersion data is available for the (spheroidal) dwarfs, 
a large
mass discrepancy will result when the conventional dynamics is used to determine the masses.
The dynamically determined mass is predicted to be larger by a factor of order 10 or more than
that which is accounted for by starsÓ \cite{milgrom83b}. This prediction was of course
ultimately born out, with dwarf galaxies having the largest mass-light ratios of any known
systems. 

The lack of linearity in Eq.~(\ref{mond_force_law}) implies that the gravitational
force on an object is not determined solely by the mass that it orbits, but 
also depends on the external gravitational field. As Milgrom
explained in detail \cite{milgrom83b,bekenstein84}, for an object like a dwarf galaxy orbiting
the Milky Way, orbits of stars within the dwarf are governed by Eq.~(\ref{mond_force_law}),
with the mass $M$ equal to the dwarf baryonic mass within the star's orbital radius,
as long as the internal gravitational acceleration $a_{\rm in}$ of the star due to the dwarf mass is
larger than the external gravitational acceleration $a_{\rm ex}$
of the dwarf as it orbits the Milky Way.
Otherwise, the dwarf is in the so-called ``quasi-Newtonian regime,''
in which case the dwarf dynamics are nearly Newtonian with an effective
gravitational constant. If the internal accelerations in the dwarf are small
compared to $a_0$, then $G\rightarrow G/\mu(a_{\rm ex}/a_0)$, and the MOND value for
the baryonic M/L ratio is simply the Newtonian value time $\mu(a_{\rm ex}/a_0)$. 
When analyzing
dwarf galaxy dynamics in the context of MOND, 
care must be taken to determine whether the gravitational
acceleration is dominated by the internal or external field. 

For the quasi-Newtonian regime, estimators for the dwarf mass based on
the stellar velocity dispersion are textbook formulas. In the opposite limit, where
the external gravitational field is negligible (the ``isolated'' regime), Eq.~(\ref{poisson_equation})
leads to the relation
$M = 9\sigma_3^4/(4Ga_0)$
for the mass $M$ of a stationary, isolated system with $a_{\rm in} \ll a_0$ \cite{milgrom94},
where $\sigma_3$ is the three-dimensional velocity dispersion of the system. We cannot
directly measure $\sigma_3$, but if the system is statistically isotropic (having the
same line-of-sight velocity dispersion $\sigma$ in all directions), then $\sigma_3 = \sqrt{3}\sigma$.
Effects which can
give significant departures from this simple mass estimator include anisotropy 
in the velocity dispersion for systems which
are not spherically symmetric and coherent velocities due to tidal effects. These are potentially
visible with sufficiently sensitive photometry and spectroscopy. Milgrom and Brada (2000)
analyzed these effects in more detail in the context of MOND, showing, for
example, that dwarfs are more susceptible to tidal disruption than for Newtonian
gravity with dark matter haloes \cite{brada00} (see also \cite{san07} for a more
detailed analysis). 

Stellar population studies strongly suggest that the baryonic mass-light ratio $\Upsilon$
for
dwarf satellites of the Milky Way should be in the range
$1 < \Upsilon < 3$ (see, e.g., \cite{bel01} and \cite{dej07}); these satellites generally have negligible gas contribution to
their baryonic mass. A total dwarf mass derived from MOND should give a value of $\Upsilon$
in this range, by comparing with the measured luminosity.
Gerhard and Spergel \cite{spergel92} and Gerhard \cite{gerhard94} analyzed seven dwarf spheroidals 
using MOND and claimed values for $\Upsilon$ ranging from around 0.2 for Fornax
to around 10 for Ursa Minor and Leo II. However, in a paper whimsically entitled
``MOND and the Seven Dwarfs'' \cite{sevendwarfs}, Milgrom persuasively
argued that published uncertainties in the measurements of velocity dispersion
used by Gerhard and Spergel were sufficient to explain the anomalous baryonic
M/L ratios. He also noted how difficult it is to determine the photometric structural
parameters of dwarfs, including central surface brightness, core radius, tidal radius,
and total luminosity. Using measurements of dwarf parameters from 
Mateo (1994) \cite{mateo94} plus several subsequent measurements, Milgrom
found that MOND was completely consistent with reasonable values for $\Upsilon$ between
1 and 3 for all seven dwarf spheroidals considered (Sculptor, Sextans, Carina, Draco,
Leo II, Ursa Minor, and Fornax). This analysis estimates the mass using only the
central velocity dispersion $\sigma$ for each dwarf, and computes $M$  in either
the quasi-Newtonian or the isolated limit. 

The most detailed analysis of the classical dwarf galaxies has been done by
Angus (2008), incorporating many improvements in observed dwarf properties over the
previous decade, and also using measured variations of the velocity
dispersion with radius in each dwarf for which it is measured \cite{angus08}. 
This analysis thus provides
a substantially more stringent test of MOND compared to the Seven Dwarfs paper. 
Angus finds best-fit values and errors
for $\Upsilon$, plus two parameters describing possible velocity dispersion anisotropy,
Of the eight dwarfs considered, six (Sculptor, Carina, Leo I, Leo II, Ursa Minor, and Fornax) are  consistent with a reasonable value of $\Upsilon$.
for the stellar content. Sextans and Draco both give discrepantly high values for 
$\Upsilon$, but within $2\sigma$ of the accepted range. Draco has been previously
known to present a
problem for MOND \cite{lokas02,lokas06}, and the problem is worse for different measurements of velocity dispersions \cite{wil04} than Angus uses. Angus notes that
distance and velocity dispersion uncertainties for these two dwarfs may have
a significant impact, and that possible tidal effects introduce uncertainties
which are large enough to explain Angus' marginally high values of $\Upsilon$ for these
two dwarfs. However, Draco in particular shows no morphological evidence of tidal
disruption. Angus speculates that it could be in the early stages of tidal heating peculiar to MOND \cite{brada00}
as it approaches the Milky Way. Numerical simulations would be required to decide
whether this explanation is plausible. An alternate explanation is that the stellar samples are
contaminated with unbound stars which are not part of the dwarf galaxy. These can inflate
the inferred line-of-sight velocities in the stellar sample, leading to systematically high
values for $\Upsilon$. (An anonymous referee for this paper claims that this is a particular issue for Sextans
and Carina, while data for Draco and Ursa Minor are not yet public.)

A provocative comparison between dark matter and MOND has recently been
performed using tidal dwarfs. Bournaud and collaborators \cite{bournaud07} have
measured rotational velocities for three gas-rich dwarfs forming from collisional debris
debris orbiting NGC5291. Using the VLA, they obtained rotation curves from the 21-cm
gas emission. The stellar mass in these systems is a small fraction of the gas mass, so
the total mass can be determined with little uncertainty. The rotation curves are symmetric,
implying that the systems are rotationally supported and in equilibrium.
It is straightforward to derive mass estimates based on Newtonian gravity: the total
mass in the systems is about three times the observed baryonic mass. The authors
claim that this is a problem based on
numerical dark matter simulations showing that gaseous dwarfs condensing out of
tidal debris should contain very little dark matter \cite{barnes92,bournaud06}. Tidal dwarfs
have complex dynamical histories and a range of baryonic effects are important, so it
is hard to know whether the comparison with simulations really is a problem. 
On the other hand, Milgrom \cite{milgrom07} and Gentile et al.\ \cite{gentile07}
have both shown that the tidal dwarfs are well accounted for with MOND,
with {\it no free parameters}. This consistency is very difficult to explain in dark matter models, as tidal
dwarfs
form from tumultuous dynamical processes with strong dependence on initial
conditions while being affected by numerous
baryonic processes. Why should their dark matter content be precisely that needed
to make them land right on the MOND force law? Clearly a larger sample of these
very interesting dynamical systems is an observational priority.

A recent interesting
set of observations has been performed on the distant and diffuse globular cluster 
Palomar 14 \cite{jordi09}, among other applications of MOND to distant
globular clusters (e.g., \cite{baumgardt05,haghi09,sollima09}). 
The radial velocities of 17 red giant stars in the cluster
 have been measured with the VLT and Keck telescopes, giving an estimate
 of the line-of-sight velocity dispersion. This dispersion is consistent with
 Newtonian gravity, even though the globular cluster should be well
 within the MOND regime, and the authors claim the observations taken at
 face value rule out MOND. However, the paper also constructs a scenario where
 the stellar velocities could still be consistent with MOND if it were on a highly
 elliptical orbit in the Milky Way potential, passing in and out of the MOND regime
 as the external gravitational field changes. This picture is related to tidal shocking 
 \cite{gnedin97,read06}
 which can decrease the stellar velocity dispersion for dwarfs with Newtonian gravity. 
 The cluster is very distant and
 faint, so prospects for obtaining a proper velocity on the sky, and thus a
 clearer characterization of the cluster's recent orbital history around the Milky
 Way, is likely remote. Mass segregation of the red giants in the cluster may also
 be an issue. But these systems put some significant pressure on MOND. 
 
The other smaller Milky Way
dwarfs which have recently been discovered in Sloan Digital Sky Survey data
(see, e.g., \cite{simon07,strigari08}) have even higher ratios of visible mass to inferred
dark matter mass than the classical
dwarf spheroidals, presenting another stringent test of MOND in galactic systems down
to still smaller scales than have been so far analyzed. Sanchez-Salcedo and Hernandez (2007)
performed a preliminary analysis concluding that the required MOND value for $\Upsilon$
increases at low luminosities and is too high to be realistic. This is a consequence of these dwarf spheroidals being consistent
with a universal central velocity dispersion. 
This provocative conclusion clearly warrants further study. Systematic effects related
to a host of assumptions about the Milky Way gravitational potential, tidal effects in the dwarfs,
stellar interlopers in the data, and other considerations can dominate the uncertainties in these
calculations.

\subsection{Can a Dwarf Galaxy Rule Out MOND?}

The results described in the previous section raise a tricky question for MOND. On one
hand, the theory is in principle more predictive than dark matter: the gravitational field
of a distribution of visible matter is completely determined given the baryonic mass-light 
ratio $\Upsilon$ and the distance, which sets the length scale in Eq.~(\ref{mond_force_law}). 
Both of these numbers can be determined or constrained by other observations.
(In practice, $\Upsilon$ is usually left as a free parameter, and then its MOND-derived value is compared to the stellar population to determine whether it is reasonable.) On the other hand, the
gravitational field is probed via the dynamics of test objects, generally the motions
of stars in the galaxy. We generally do not have available the full velocity vector or velocity dispersion, but only the line-of sight-components. The transverse components
contain crucial information about the full dynamical state of the dwarf: did its
past trajectory create tidal heating or disruption? To what extent is its internal velocity
dispersion anisotropic? How rapidly is the external gravitational field varying? Without this
information, a range of gravitational potentials are consistent with the current dynamical
state of any galaxy. In the absence of other difficult observations (like proper motions, which
allow reconstructing the actual galaxy trajectory), we must settle for determining whether the observed galaxy dynamics requires only believable additional assumptions about unobservable effects. 

This is the same issue facing MOND fits of large galaxies as well. Large galaxies
tend to have much less significant external gravitational field effects than dwarfs, leaving
less freedom to explaining marginal MOND fits. The remarkable point about MOND is
that, despite the limited freedom the model provides in fitting galaxy dynamics, 
very few galaxies, ranging over several orders of magnitude in mass and more than
an order of magnitude in size, provide a true challenge for MOND.

In contrast, the dark matter situation is more forgiving. The standard cosmology in principle
is completely deterministic, once the properties of dark matter are specified. Small initial
density perturbations, specified by measurements of the cosmic microwave background
temperature anisotropies, evolve according to completely specified equations of 
cosmological evolution. However, we are not yet able to make detailed  {\it ab initio} predictions for
the dark matter halo associated with a particular class of galaxies. Since the baryonic
mass is coupled gravitationally to the dark matter, a complete understanding of
the dark matter distribution today requires detailed knowledge of the history
of the baryon distribution, and this in turn is shaped by complex
baryonic physics, including heating, cooling, star formation and evolution, and energy feedback from stars and quasars. This is not simply an academic point: most of the outstanding issues with dark matter involve small scales, where the baryon densities can be comparable to the dark matter densities. Eventually, our modeling and computations may reach the point where
we can simulate the evolution of realistic galaxies starting only from cosmological initial conditions,
but this prospect is many years away. Until then,
visible galaxies' rotation curves or velocity dispersions are generally fit using a parameterized
model for the dark matter halo, and we have few constraints on the halo parameters for an
individual galaxy. So for an individual galaxy, fitting a dark matter halo provides more freedom
than fitting a MOND value for $\Upsilon$.

\subsection{Dwarfs as Dynamical Tracers}

The motions of dwarf galaxies themselves as they orbit around their
parent galaxy can also be used as a test of MOND. Typically, for large
elliptical galaxies, satellite galaxies are observable to much larger radii
than the stars in the central galaxy, so the satellites probe the
gravitational field over a larger region. However, any individual galaxy
has a relatively small number of observable satellites, making gravitational
potential constraints inconclusive. To get around this, Klypin and Prada \cite{klypin09}
have used a sample of 215,000 red galaxies with redshifts between $z=0.010$ and $z=0.083$
from the Sloan Digital Sky Survey Data Release 4. From this catalog, they select
a set of satellites according to the criteria that the satellite is at least a factor of
four fainter than its host, has a projected distance from its host of less than 1 Mpc, and
has a velocity difference from its host of less than 1500 km/s. This results in
about 9500 satellites around 6000 host galaxies. They then form a composite
satellite velocity dispersion as a function of radius for a number of bins in
host galaxy luminosity by stacking all of the satellites and
using a maximum-likelihood estimator. 
A homogeneous background level, reflecting random alignments of unrelated
galaxies, is subtracted to give a final velocity dispersion for each luminosity bin. 

Klypin and Prada compare this velocity dispersion with circular dark matter velocities
around large isolated haloes drawn from N-body simulations, and find very good
agreement. They make the assumption that the density of dwarf galaxies traces
the dark matter density, which holds in dark matter simulations \cite{sales07}. They further claim that velocity dispersions derived by solving the
Jeans equation with the MOND gravitational potential
in Eq.~(\ref{poisson_equation}) can only fit the data using highly contrived
forms for the velocity dispersion anisotropy, corresponding to highly elliptical stellar orbits. 
This conclusion has been disputed by Angus et al.\ (2008) \cite{angusetal08}, who claim that the
same anisotropy is actually very reasonable, based on numerical
simulations of elliptical galaxy formation in MOND \cite{nipoti07}. The two papers also
disagree on a number of other points, including whether gravitational
fields external to the SDSS galaxies should be relevant and what constitutes reasonable mass-light
ratios for the host galaxies. Further independent analysis 
of this interesting data set is clearly warranted.

\section{Galaxy Clusters and MOND}

While the Milgrom force law, Eq.~(\ref{mond_force_law}), and its generalization
to the modified Poisson equation in Eq.~(\ref{poisson_equation}), is a highly
successful phenomenological description of galaxies ranging from the smallest dwarfs
to the largest spirals, and over the full range of observed surface brightness, its efficacy
does not extend to the larger scales of galaxy clusters.
A typical galaxy cluster must contain a total mass several
times its visible gas and stellar mass for its gas temperatures and galaxy velocities 
to be explained by the Milgrom
force law \cite{aguirre01}. Simply postulating an increased cluster value for the baryonic mass-light 
ratio $\Upsilon$ is not realistic, since the baryonic mass is dominated by gas and does not
have the intrinsic uncertainty in $\Upsilon$ that stellar populations do. In the absence of large systematic errors in the cluster data, we have several choices for explaining galaxy clusters: 

\begin{enumerate}

\item
MOND is simply not a fundamental theory, but only a phenomenological relationship for the distribution of dark matter
in galaxies, inapplicable to larger bound systems. 

\item
MOND does represent a fundamental force law, but clusters contain additional dark matter while galaxies do not. 

\item
MOND represents a fundamental
force law for galaxies, but this force law gets modified on larger scales. 

\end{enumerate}

\noindent
The first possibility is widely accepted: galaxy clusters simply rule out MOND as a modification of gravity. The second possibility is not as outlandish as it appears at first glance. If dark matter were in the form of light neutrinos with a mass of around 2 eV, for example, which had relativistic velocities at the time of their decoupling, they would not be gravitationally bound in galaxies, but could be bound in the much larger and more massive clusters \cite{sanders03,angus07}. Current best limits on neutrino masses come from cosmology, but
these assume the standard cosmological model, which may be modified if MOND actually represents
a fundamental force law. Angus \cite{ang09} postulates a slightly heavier (11 eV) sterile neutrino, which
a numerical calculation shows can reproduce the acoustic peaks in the microwave background assuming that early-universe physics is not affected by MOND effects. Producing the forced acoustic oscillations
revealed by precision measurements of the microwave background power spectrum is an important
challenge for any cosmology with different dark matter or gravitation from the standard cosmology. Dynamical
analysis shows that such a neutrino can reproduce observed galaxy cluster phenomenology while
not affecting galaxy rotation curves \cite{ang09b}. 
Pressure is put on these models from analyses of gravitational
lensing \cite{nat08,fer08,tian09}, although these all make specific assumptions about lensing
in MOND, which is not uniquely determined without the context of a relativistic theory.



The most relevant observations for MOND are the gravitational lensing observations of the famous
``bullet'' cluster 1E 0657-558 \cite{clowe06} and the cluster MACS J0025.4-1222 \cite{bradac09}. 
In these systems of two merging galaxy clusters, the
peak of the gravitational potential, revealed by weak lensing, is offset from the peak of the
mass distribution, visible in hot gas. This offset is consistent with the clusters containing substantial
dark matter; when the clusters merge, the dark matter, which is assumed to be collisionless, 
continues to move unimpeded, while the gas composing the bulk of the visible matter is slowed
via shock heating. The offset is NOT consistent with MOND, or with {\it any} modified gravity theory where
the gravitational field is generated solely by the visible matter: the geometry of the gravitational
field is different from the geometry of the visible matter \cite{sellwood02}. The merging-cluster 
observations decisively
demonstrate that Eq.~(\ref{poisson_equation}) cannot by itself describe gravitation. This is
consistent with the earlier observations showing the inferred cluster dark matter content is discrepant with
the MOND force law prediction, but further rules out the discrepancy being due to additional modifications
of the gravitational force law on cluster scales (case 3 above). Unlike individual galaxies, galaxy clusters 
{\it must} possess some source of gravitation in addition to that provided by the visible matter (i.e., dark matter!). 

\section{Relativistic MOND and Dark Matter}

Many people felt relieved that the cluster lensing results meant they could finally
stop hearing about all of this pesky MOND business. However, option 2
above is still viable, although most cosmologist view it as an unnaturally complicated
theoretical possibility. As mentioned above, clusters could contain hot dark matter,
like light neutrinos, which would be sufficient to explain the observed gravitational
and gas morphology of the merging clusters. While this solution invokes two separate
pieces of new physics (sufficiently massive neutrinos and a MOND force law), it is
not immediately obvious how much more speculative this should be considered
than standard dark matter models, given that neutrinos are already
known to have non-zero masses from neutrino oscillation observations. 
But the most interesting
option for MOND may come from a different direction. 

Two years before the bullet cluster observations showed that the gravitational field must be
generated by more than just the visible matter, Jacob Bekenstein published a relativistic gravitation
theory which reduced to MOND in the point-source limit \cite{bekenstein04}. 
For many years, the fact that MOND had
no relativistic basis was often cited as a compelling argument against the idea.
(I sometimes joked that people claimed a relativistic MOND extension was ruled out by the Bekenstein
Theorem. This theorem stated that Bekenstein had tried to do it and failed,  
so therefore it was not possible. Later, Bekenstein
told me that he had actually never put much effort into the problem.) With a full relativistic
theory of gravitation, diverse observations ranging from solar system constraints on
post-Newtonian parameters to gravitational lensing, cosmological structure formation,
and the expansion history of the universe, can be addressed. All of these areas were speculated about using the MOND force law or modified Poisson equation, and reasoning by
analogy with general relativity. But this is clearly an unsatisfactory approach: any results contradicting observations can simply be explained away as the result of an insufficiently
sophisticated theory.

Several variations on
the Bekenstein proposal are now on the table \cite{sanders05,navarro06}. 
These theories all share the common element
of additional dynamical fields which act as gravitational sources. The most elegant
scheme so far has been dubbed Einstein Aether gravity \cite{zlosnik07}. A dynamical
time-like vector field $A_\mu$ with $A_\mu A^\mu = -1$ 
is incorporated into gravity by adding a term ${\cal F}({\cal K})$
to the Einstein-Hilbert gravitational action, where ${\cal F}$ is a function to be determined and
\begin{equation}
{\cal K} \propto c_1\nabla_\mu A_\nu\nabla^\mu A^\nu
+ c_2 (\nabla_\mu A^\mu)^2 + c_3\nabla_\mu A^\nu \nabla_\nu A^\mu
\label{vector_term}
\end{equation}
with $c_1$, $c_2$, and $c_3$ undetermined constants. This is an unusual theory because
the time-like vector field picks out a preferred frame at each point in spacetime (in which the time
coordinate direction aligns with $A$) and thus violates local Lorentz invariance. But if this
symmetry breaking is small enough it would be below the level of current experimental bounds. 
This class of theories has a line of theoretical precedents (see, e.g., \cite{will73,jacobson01,carroll04}).

Remarkably, the constants in ${\cal K}$ and the function ${\cal F}$ can be chosen so that the nonrelativistic limit
of the theory reduces to a modified Poisson equation of the form Eq.~(\ref{poisson_equation}):
the MOND limit can be reproduced in this class of relativistic models. Much analysis has yet to
be performed, but it is at least plausible that the excess gravitational forces induced by
the vector field $A^\mu$ could match the departures from MOND seen in galaxy clusters. Relativistic MOND theories involving extra dynamical fields are, in some sense,
an abdication of the original MOND philosophy that the visible matter creates all of the
observed gravitational force in the universe; but we are forced into this by observations.
The physical content of these theories is 
quite different from traditional  models; the ``dark sector'' is governed by 
physics different from particle dark matter.

\section{Challenges}

Relativistic extensions of MOND have not been widely explored, in contrast to
the overwhelming amount of work on the standard cosmological model. So it is simply
not yet known whether particular relativistic MOND theories can reproduce all
known gravitational phenomena, including gravitational lensing, post-Newtonian
constraints from the solar system and binary pulsars, tidal streams, 
galaxy cluster gravitational potentials,
and the observed dynamics of merging galaxy clusters. All of these phenomena are
successfully explained by the standard dark matter cosmology, and represent a tall
order for any new gravitational theory. Relativistic MOND theories also must be nonlinear,
since MOND itself is, making these analyses challenging. However, relativistic
MOND theories hold the promise of explaining two observations that the standard cosmology
cannot: (1) the effectiveness of the Milgrom force law, Eq.~(\ref{mond_force_law}), and
related phenomenology of galaxy dynamics, and
remarkably, (2) the accelerating expansion of the Universe. 

Milgrom immediately realized that 
the MOND acceleration scale $a_0$ is the same order as $cH_0$ and
thus also the same order as $c\Lambda^{1/2}$, where $H_0$ is the Hubble
parameter and $\Lambda$ is the cosmological constant which would explain
the accelerating expansion of the universe. 
It is difficult to see how $a_0$, which manifests itself on scales of dwarf galaxies,
can be related to either $H_0$ or $\Lambda$ 
in the standard cosmological
model, but a relativistic basis for MOND may provide insight. 
Ref.~\cite{zlosnik07} discusses conditions under which the additional vector
field in Einstein Aether gravity
will influence the expansion rate of the Universe. In fact, they display a
choice of ${\cal F}$ which gives a late-time expansion history identical to that of
general relativity with a cosmological constant. This is highly provocative: could
a conceptually simple modification of relativistic gravitation explain the phenomenology
of both dark matter and dark energy? The particular theory considered in \cite{zlosnik07}
has substantial freedom in it, and is also challenging to analyze because it is
nonlinear. But substantial effort in this direction is clearly warranted, as the
theory easily reproduces MOND-like behavior and accelerating cosmological expansion
with different choices of the free parameters and the free function. 

Most cosmologists pay little attention
to the highly successful galaxy phenomenology of MOND, 
even though they should, because they do not believe
in MOND as a fundamental theory.  The merging galaxy cluster observations have now
ruled out the possibility that Eq.~(\ref{poisson_equation}) applied to visible matter represents the
entire fundamental theory of gravitation, 
while dark matter cosmology has had remarkable success in
explaining the growth of structure in the Universe and providing a simple standard cosmological
model which explains most observations. On the other hand, dark matter cosmology still
faces some serious issues, mostly related to reproducing this MOND phenomenology
(with the obvious exception of dark energy). 
The recent observation implying  tidal dwarfs obey MOND \cite{gentile07}
is one intriguing example. We should not allow the successes of our leading
theories to blind us to other possibilities. 

The discovery of the accelerating expansion of the Universe changed the playing field for
cosmology. Prior to the late 1990's, many established cosmologists would greet any suggestion
of a modification of gravity as the explanation for observed gravitational force excesses with
contempt. Today, some of these same scientists author papers advocating a modification
of gravity as a possible explanation for the observed accelerating expansion. Until the predominant picture of dark matter cosmology can explain all of the observations, other competing ideas should be pursued, either as a way of sharpening the case for dark matter cosmology, or, perhaps, uncovering an eventual replacement. 

\acknowledgments  I thank Andrew Zentner for insightful discussions about galaxy formation. Two
anonymous referees helpfully pointed out recent relevant literature, and one emphasized how unbound stellar interlopers can increase inferred mass-light ratios in dwarf spheroidals.
This work was partially supported by NSF grant AST-0807790.

\end{document}